\documentclass[journal]{new-aiaa}
\usepackage[utf8]{inputenc}
\usepackage{textcomp}
\usepackage{graphicx}
\usepackage{amsmath}
\usepackage[version=4]{mhchem}
\usepackage{siunitx}
\usepackage{longtable,tabularx}
\usepackage[capitalise]{cleveref}
\usepackage{subcaption}
\title{Optimal Rank-1 Directional State Transition Tensors}

\author{Grace E. Calkins\footnote{Ph.D. Student, Department of Aerospace Engineering Sciences. grace.calkins@colorado.edu} and Jay W. McMahon,\footnote{Associate Professor, Department of Aerospace Engineering Sciences.}}
\affil{University of Colorado Boulder, Boulder, CO 80303}

\author{Jackson Kulik \footnote{Assistant Professor, Mechanical and Aerospace Engineering}}
\affil{Utah State University, Logan, Utah, 84322}

\begin{document}

\maketitle

\section{Introduction}
Nonlinear uncertainty quantification schemes are required for aerospace guidance, navigation, and control tasks under high state uncertainty and dynamical nonlinearity. Methods to address these challenges are increasingly important to solve problems such as tracking of the large number of satellites in Cislunar space or navigation during aerocapture on Uranus. State transition tensors (STTs) and differential algebra methodologies have been employed to address these problems \cite{park2006nonlinear,armellin_asteroid_2010}. However, efficiency degrades as Taylor series expansion order increases, since the polynomial order of the computational complexity of most algorithms varies linearly or superlinearly in the highest order of the Taylor series. To increase computational efficiency, Boone and McMahon developed the directional state transition tensor (DSTT) approximation \cite{boone2023directional} by making the fortuitous observation that in many instances the dominant effects of the state transition tensors are largely captured by just their action on the dominant right singular subspace of the state transition matrix. In their work, Boone and McMahon showed that the storage complexity of DSTTs can be only fractionally larger than the storage of the state transition matrix itself, and that statistical moment propagation can be accomplished approximately with the DSTTs in fractions of the time of the full moment propagation by STTs \cite{park2006nonlinear, acciarini2025nonlinear}. 

Boone and McMahon have explored the use of DSTTs in approximating the second-order Gaussian extended Kalman filter with just slightly more computational cost than the original extended Kalman filter \cite{boone2024efficient}. However, Calkins et al. have shown that the original approach to DSTT computation may not always yield good approximations to the STTs, especially in the case of aerocapture \cite{calkins2024efficient,calkins2025dynamics}. Additional work has been conducted to improve the computational efficiency of DSTT computation and obtain more accurate DSTTs for particular applications. Zhou et al. have developed methods to compute the DSTT without having to first calculate the state transition matrix (STM) or the entire STT \cite{zhou2024time}. Zhou et al. have also developed DSTTs based on a different choice of ``important'' factors besides the dominant right singular subspace of the STM, employing the subspace that is most important to the particular orbit determination application \cite{zhou2025efficient}. 

In parallel to the development of DSTTs, related work on tensor eigenvalues, induced norms, and nonlinearity indices associated with constrained optimization problems involving the STTs and related quantities has progressed. Jenson and Scheeres developed nonlinearity metrics using STTs \cite{jenson2023semianalytical, jenson2024bounding}, and Kulik et al. have applied tensor norms to guidance, navigation, and control, Gaussian mixture model composition, and linear covariance propagation validation \cite{kulik2024applications, kulik2024nonlinearity, kulik2025linear}. These approaches are pertinent to DSTTs through the connection between the best rank-1 approximation of a supersymmetric tensor and the maximal tensor z-eigenvalue and eigenvector pairs. By deriving and computing optimal rank-1 directional state transition tensors, we hope to better understand the original DSTT approach and the conditions under which the dominant subspace of the state transition matrix yields nearly optimal approximations.

This note is organized as follows. First, a mathematical background for the tensor algebra germane to this paper is presented. Then, the optimal partially-symmetric rank-1 approximation of a $(1,m)$-tensor is derived. Next, the optimal rank-1 DSTT approximation is compared with the original DSTT approach for uncertainty quantification. Finally, applications are presented, followed by concluding remarks. 

\section{Background}

This work will make use of a number of tensor operations outlined here. The Frobenius inner product between two tensors $\mathbf{A}\in\mathbb{R}^{n_1 \times n_2 \times \ldots \times n_m}$ and $\mathbf{B}\in\mathbb{R}^{n_1 \times n_2 \times \ldots \times n_m}$ is the inner product between the tensors when flattened into vectors \cite{lim2013tensors},
\begin{equation} \label{eqn:fro_inner_prod}
    \langle \mathbf{A},\mathbf{B}\rangle_F = \bar{a}_{i;j_1 \ldots j_m} \bar{b}_{i;j_1 \ldots j_m},
\end{equation}
where $\bar{a}_{i;j_1 \ldots j_m}$ and $\bar{b}_{i;j_1 \ldots j_m}$ are the entries of $\mathbf{A}$ and $\mathbf{B}$ when flattened. Einstein notation is used, where subscripts indicate the elements of the tensor, repeated indices indicate summation, and semicolons separate the input and output indices. 
The Frobenius inner product induces the Frobenius norm of a real-valued $m^{\text{th}}$-order tensor $\mathbf{A}\in\mathbb{R}^{n_1 \times n_2 \times \ldots \times n_m}$ with entries $a_{i_1 i_2 \ldots i_m}$ under the Euclidean metric is the square root of the sum of the squared entries:
\begin{equation}
    \|\mathbf{A}\|_F^2 = \sum_{i_1=1}^{n_1} \sum_{i_2=1}^{n_2} \cdots \sum_{i_m=1}^{n_m} |a_{i_1 i_2 \ldots i_m}|^2.
\end{equation}
The tensor product of three vectors $\mathbf{u}, \mathbf{v}, \mathbf{w} \in \mathbb{R}^n$ is defined by
\begin{equation}
    (\mathbf{u} \otimes \mathbf{v} \otimes \mathbf{w})_{ijk} = u_i v_j w_k,
\end{equation}
which results in a tensor in $\mathbb{R}^{n\times n \times n}$. 

The main tensor under consideration in this work is he $m^{\text{th}}$-order STT, denoted here simply as $\boldsymbol{\Phi}$, is the coefficient tensor appearing in the Taylor series expansion of the flow of a dynamical system. The STT is a $(1,m)$-tensor, meaning that it is symmetric under permutations of the latter $m$ indices and can be viewed as taking $m$ input vectors (or $m$ copies of the same input vector) and transforming that into a vector output. 

To form DSTTs, Boone and McMahon showed that the STTs can be rotated such that the partials are taken with respect to $\mathbf{y} \in \mathbb{R}^n$ so long as $\mathbf{y}$ constitutes an orthogonal basis that spans $\mathbb{R}^n$ \cite{boone2023directional}. To reduce the dimension of the DSTTs, the dimension of basis $\mathbf{y}$ can be reduced to $\mathbb{R}^k$ such that $k < n$. Choosing $k=1$, such that $y$ is a scalar, produces a rank-1 DSTT (referred to in this work as a R1-DSTT); however, a larger $k$ can be chosen to construct a rank-$k$ DSTT. The second- and third-order DSTTs, $\boldsymbol{\psi}^{[2]}$ and $\boldsymbol{\psi}^{[3]}$, are obtained from the second- and third-order STTs $\boldsymbol{\Phi}$ through the change of basis
\begin{align} \label{eqn:stt2dstt}
    \psi_{i;\gamma_1 \gamma_2}^{[2]} &= \phi_{i;\kappa_1 \kappa_2} R_{\gamma_1; \kappa_1} R_{\gamma_2; \kappa_2}, \quad \text{and} \\
    \psi_{i;\gamma_1 \gamma_2 \gamma_3}^{[3]} &= \phi_{i;\kappa_1 \kappa_2 \kappa_3} R_{\gamma_1; \kappa_1} R_{\gamma_2; \kappa_2} R_{\gamma_3; \kappa_3},
\end{align}
where $\mathbf{R} \in \mathbb{R}^{k\times n}$. Time indices are omitted from the STTs and DSTTs in the following sections for brevity.

\section{Optimal Partially-Symmetric Rank-1 Approximation of a $(1,m)$-Tensor}
\label{sec:ODSTTs}
We treat the problem of developing the best partially-symmetric rank-1 approximation of a $(m+1)$-order tensor such as the $m^{\text{th}}$-order STT $\boldsymbol{\Phi}$ in the Frobenius-norm, following a generalization of the procedure in \cite{qi2006rank}, by solving the optimization problem
\begin{equation} \label{eqn:optimization}
\mathbf{u},\mathbf{v}=\arg\min_{\mathbf{y},\Vert\mathbf{x}\Vert_2=1}\Vert \boldsymbol{\Phi}-\mathbf{y}\otimes\mathbf{x}\ldots\underset{m\text{ times}}{\otimes}\ldots\mathbf{x} \Vert_F^2,
\end{equation}
in order to find the optimal rank-1 partially symmetric approximation
\begin{equation}
    \boldsymbol{\Phi}\approx\mathbf{u}\otimes\mathbf{v}\ldots\underset{m\text{ times}}{\otimes}\ldots\mathbf{v}.
\end{equation}
To find these optimal factors $\mathbf{u}$ and $\mathbf{v}$, the squared Frobenius norm on the right-hand side of \cref{eqn:optimization} is expanded as
\begin{equation} \label{eqn:optimization2}
    \| \boldsymbol{\Phi} \|_F^2 + \|\mathbf{y}\otimes\mathbf{x}\ldots\underset{m\text{ times}}{\otimes}\ldots\mathbf{x}\|_F^2 - 2 \langle \boldsymbol{\Phi},\mathbf{y}\otimes\mathbf{x}\ldots\underset{m\text{ times}}{\otimes}\ldots\mathbf{x}\rangle_F,
\end{equation}
and applying the unit-norm constraint on $\mathbf{x}$, the second term of \cref{eqn:optimization2} is:
\begin{equation}    
    \|\mathbf{y}\otimes\mathbf{x}\ldots\underset{m\text{ times}}{\otimes}\ldots\mathbf{x}\|_F^2 = \| \mathbf{y}\|^2_F \| \mathbf{x}\|_F^2 \underset{m\text{ times}}{\ldots} \| \mathbf{x}\|_F^2  = \| \mathbf{y}\|^2_F.
\end{equation}
This constraint is reasonable since $\mathbf{y}$ can be scaled up or down to achieve the same effect as varying the scale of $\mathbf{x}$. By invoking this simplification and the definition of the Frobenius norm inner product in \cref{eqn:fro_inner_prod}, the optimization problem from \cref{eqn:optimization} becomes
\begin{equation} \label{eqn:optimization_expanded}
    \mathbf{u},\mathbf{v}=\arg\min_{\mathbf{y},\Vert\mathbf{x}\Vert_2=1}\Vert \boldsymbol{\Phi} \Vert_F^2 - 2\Phi_{i;j_1...j_m}y_i x_{j_1}\ldots x_{j_m}+\Vert\mathbf{y}\Vert_2^2.
\end{equation}
Let the vector $\mathbf{c}$ be defined such that
\begin{equation}
    c_i=\Phi_{i;j_1...j_m} x_{j_1}\ldots x_{j_m}.
\end{equation}
Then, the optimization in \cref{eqn:optimization_expanded} is equivalent to
\begin{equation}   
    \mathbf{u},\mathbf{v}=\arg\min_{\mathbf{y},\Vert\mathbf{x}\Vert_2=1}\Vert \boldsymbol{\Phi} \Vert_F^2 - 2\mathbf{c}^T\mathbf{y}+\Vert\mathbf{y}\Vert_2^2.
\end{equation}
This expression is quadratic in $\mathbf{y}$, and is minimized by $\mathbf{u}=\mathbf{y}^*=\mathbf{c}$. The optimization is then reduced to
\begin{align}
    \mathbf{v}&=\arg\min_{\Vert\mathbf{x}\Vert_2=1}\Vert \boldsymbol{\Phi} \Vert_F^2 -\Vert\mathbf{c}\Vert_2^2\\
    &=\arg\min_{\Vert\mathbf{x}\Vert_2=1}\Vert \boldsymbol{\Phi} \Vert_F^2 -\Phi_{i;j_1...j_m} x_{j_1}\ldots x_{j_m}\Phi_{i;k_1...k_m} x_{k_1}\ldots x_{k_m},
\end{align}
which is equivalent to finding
\begin{equation}
\label{eq:var_eig}
\arg\max_{\Vert\mathbf{x}\Vert_2=1}\Vert\boldsymbol{\Phi}\mathbf{x}^m\Vert_2^2,
\end{equation}
where $\boldsymbol{\Phi}\mathbf{x}^m$ denotes the contraction of all input indices with copies of the vector $\mathbf{x}$
\begin{equation}
(\boldsymbol{\Phi}\mathbf{x}^m)_i=\Phi_{i;j_1,...,j_m}x_{j_1}...x_{j_m}.
\end{equation}
The solution to this problem is given by the z-eigenvector corresponding to the maximal z-eigenvalue of the ``square'' $\tilde{\boldsymbol{\Phi}}$ of the tensor $\boldsymbol{\Phi}$, where
\begin{equation}
    \tilde{\Phi}_{j_1\ldots j_m,k_1\ldots k_m}=\Phi_{i;j_1...j_m}\Phi_{i;k_1...k_m}.
\end{equation}
The z-eigenvalue equation is found by applying the method of Lagrange multipliers to \cref{eq:var_eig} to obtain
\begin{equation}
    \bar{\boldsymbol{\Phi}}\mathbf{x}^{2m-1}=\lambda\mathbf{x}, \quad \Vert\mathbf{x}\Vert_2=1,
\end{equation}
where
\begin{equation}
    (\bar{\boldsymbol{\Phi}}\mathbf{x}^{2m-1})_{j_1}=\bar{\Phi}_{j_1\ldots j_m,k_1\ldots k_m}x_{j_2}\ldots x_{k_{2m}},
\end{equation}
and $\bar{\boldsymbol{\Phi}}=\mathrm{sym}(\tilde{\boldsymbol{\Phi}})$ is the symmetrization or average over all permutations of the indices. The maximal z-eigenvector of the ``square'' of the state transition tensor can be found by employing shifted symmetric higher-order power iteration (SS-HOPM) \cite{kolda2011shifted}. This algorithm can be augmented as we are solving for the eigenpairs of a ``squared'' tensor. The symmetrization step can be avoided following the appendix of \cite{kulik2024applications}, and a less conservative shift factor than in \cite{kolda2011shifted} can be computed following the appendix of \cite{siciliano2025higher}.

To summarize, the best rank-1 partially symmetric approximation of the $(m+1)$-order state transition tensor (hereafter referred to as an optimal rank-1 directional state transition tensor or R1-ODSTT) under the Frobenius-norm is given by $\mathbf{u}\otimes\mathbf{v}\ldots\underset{m\text{ times}}{\otimes}\ldots\mathbf{v}$ where $\mathbf{v}$ is given by the z-eigenvector corresponding to the maximal z-eigenvalue of the ``square'' of the state transition tensor, and $\mathbf{u}=\boldsymbol{\Phi}\mathbf{v}^m$ is given by operating the state transition tensor on the vector $\mathbf{v}$. In addition to being optimal in the Frobenius-norm sense, this R1-DSTT construction simplifies online computations, as perturbation propagation reduces to matrix and vector multiplication instead of tensor contraction.

The error incurred when using the R1-ODSTT to propagate vectors on the unit ball is formalized as follows. The squared Frobenius norm of the approximation error is given simply by
\begin{align}
    \Vert\boldsymbol{\Phi}-\mathbf{u}\otimes\mathbf{v}\ldots\underset{m\text{ times}}{\otimes}\ldots\mathbf{v}\Vert_F^2&=\Vert\boldsymbol{\Phi}\Vert_F^2-\Vert\mathbf{u}\Vert_2^2\\
    &=\Vert\boldsymbol{\Phi}\Vert_F^2-\Vert\boldsymbol{\Phi}\Vert_2^2,
\end{align}
where $\Vert\mathbf{u}\Vert_2$ corresponds to the maximum magnitude of the output of the state transition tensor when inputs are constrained to the unit sphere, which is equivalent to the squared induced 2-norm of the state transition tensor $\Vert\boldsymbol{\Phi}\Vert_2$ \cite{kulik2024applications}:
\begin{equation}
    \Vert\boldsymbol{\Phi}\Vert_2=\max_{\Vert\mathbf{x}\Vert_2=1}\Vert \boldsymbol{\Phi}\mathbf{x}^{m}\Vert_2.
\end{equation}
Since the Frobenius norm is always positive, this leads to the inequality
\begin{equation}
\Vert\boldsymbol{\Phi}\Vert_F^2\geq\Vert\boldsymbol{\Phi}\Vert_2^2,
\end{equation}
for any $(1,m)$-tensor, which implies that the Frobenius norm of the approximation error always dominates the induced 2-norm of the approximation error. Thus, the largest possible error when employing the R1-ODSTT to propagate perturbation vectors on the unit ball is smaller than the Frobenius norm of the error:
\begin{equation} \label{eqn:fro_norm_of_error}
    \Vert\boldsymbol{\Phi}-\mathbf{u}\otimes\mathbf{v}\ldots\underset{m\text{ times}}{\otimes}\ldots\mathbf{v}\Vert_F^2\geq\Vert\boldsymbol{\Phi}-\mathbf{u}\otimes\mathbf{v}\ldots\underset{m\text{ times}}{\otimes}\ldots\mathbf{v}\Vert_2^2.
\end{equation}

\section{Comparison of R1-ODSTTs and R1-DSTTs}

For R1-DSTTs, the second-order R1-ODSTT $\mathbf{u}$ and $\mathbf{v}$ are equivalent (in form and function but not necessarily value) to the reduced-dimension R1-DSTT $\boldsymbol{\psi}^{[2]}$ and $\mathbf{R}$ matrices presented by Boone and McMahon \cite{boone2023directional}. The two differences between a R1-DSTT and a R1-ODSTT are the procedure to obtain the input and output directions and that the input directions are not necessarily the same between orders. The R1-DSTT $\mathbf{R}$ direction is the same between all orders by construction, i.e. $\mathbf{v}^{[2]} \neq \mathbf{v}^{[3]}$, whereas $\mathbf{v}^{[m]}$ is the $m^{\text{th}}$-order R1-ODSTT input direction. The simplified deterministic perturbation propagation equation and moment propagation equations developed for the R1-DSTTs (for scalar $y=\mathbf{R}^{\top} \mathbf{x}$) can also be used with R1-ODSTTs by projecting the state along the R1-ODSTT input direction, $y = \left(\mathbf{v}^{[m]}\right)^{\top} \mathbf{x}$. 

A benefit of the original R1-DSTT formulation for scalar ${y}$ is simplification to vector products instead of tensor products in deterministic perturbation and Gaussian moment propagation (for initial mean perturbations of zero). The full STT deterministic perturbation propagation equation is \cite{park2006nonlinear}:
\begin{equation} \label{eqn:stt_pert}
    \delta x(t_k)_i = \sum_{p=1}^M \frac{1}{p!} \phi_{i;\gamma_1 \ldots \gamma_p} \delta x(t_0)_{\gamma_1} \ldots \delta x(t_0)_{\gamma_p},
\end{equation}
and the full STT mean and covariance propagation equations for expansion order $M$ are \cite{park2006nonlinear}:
\begin{align} \label{eqn:stt_mean}
    \delta m(t_k)_i &= \sum_{p=1}^M \frac{1}{p!} \phi_{i;\gamma_1 \ldots \gamma_p} \mathbb{E}\left[ \delta x(t_0)_{\gamma_1} \ldots \delta x(t_0)_{\gamma_p} \right], \qquad \text{and} \\  \label{eqn:sttcov}
    P(t_k)_{ij}&= \Biggl( \sum_{p=1}^M \sum_{q=1}^M \frac{1}{p!q!} \phi_{i;\gamma_1 \ldots \gamma_p} \phi_{j;\kappa_1 \ldots \kappa_q} \nonumber  \\
    &  \quad \times \mathbb{E}\left[ \delta x(t_0)_{\gamma_1} \ldots \delta x(t_0)_{\gamma_p} \delta x(t_0)_{\kappa_1} \ldots \delta x(t_0)_{\kappa_q} \right]  \Biggl) \\
    & \quad - \delta m(t_k)_i \delta m(t_k)_j, \nonumber 
\end{align}
where the $(t_k, t_0)$ time mapping indices are omitted from the STTs (and R1-DSTTs and R1-ODSTTs in the following expressions) for clarity. When propagating a perturbation with R1-ODSTTs, keeping the full STM for full linear accuracy, \cref{eqn:stt_pert} becomes: 
\begin{equation} \label{eqn:dstt_deterministic}
    \delta x(t_k)_i = \phi_{i;\gamma_1} \delta x(t_0)_{\gamma_1} +  \sum_{p=2}^M \frac{1}{p!} u^{[p]}_{i} \left(\delta y^{[p]}(t_0)_{\gamma_1}\right)^p,
\end{equation}
where $\delta y^{[p]}(t_0) = \left(\mathbf{v}^{[p]}\right)^{\top} \delta \mathbf{x}(t_0)$. Instead of requiring tensor contraction for the higher-order terms, R1-ODSTT higher-order contributions are computed with only dot products and scalar-vector multiplication.  

One of the more costly parts of computing the quantities in \cref{eqn:stt_mean} and \cref{eqn:sttcov} is evaluating the higher-order moments of $\delta \mathbf{x}$, despite the odd-order moments reducing to zero for $\delta \mathbf{m}(t_0) = 0$. When using a R1-DSTT or R1-ODSTT, instead of propagating the moments of $\mathbf{x}$, we propagate the moments of the scalar ${y} = \mathbf{R}^{\top} \mathbf{x}$ or ${y} =\left(\mathbf{v}^{[m]}\right)^{\top} \mathbf{x}$. Because $\mathbf{R}$ and $\mathbf{v}^{[m]}$ are deterministic, they can be brought inside the expectation and to reduce the higher-order tensor contractions to scalar and vector multiplication. To simplify moment propagation, we can use the projection of the variance along the input direction $\mathbf{v}^{[m]}$:
\begin{equation} \label{eqn:sigma_r}
    \sigma_{v^{[m]}}(t_0) = \sqrt{v^{[m]}_i P(t_0)_{ij} v^{[m]}_j}. 
\end{equation}

When keeping the full STM in the expansion for mean and covariance propagation, this results in the simple third-order propagation equations for mean and covariance: 
\begin{align}
    \delta m(t_k)_i &= \frac{1}{2} u^{[2]}_i \sigma_{v^{[2]}}(t_0)^2, \qquad \text{and} \label{eqn:dsttmean} \\[6pt]
    P(t_k)_{ij} &= \phi_{i;\gamma_1} \phi_{j;\gamma_2} P(t_0)_{\gamma_1 \gamma_2} \nonumber \\
    &\quad + \frac{1}{2} u^{[2]}_i u^{[2]}_j \sigma_{v^{[2]}}^4  
    + \frac{1}{2} \!\left[\phi_{i;\gamma_1} u^{[3]}_j P(t_0)_{\gamma_1 \kappa_1} v^{[3]}_{\kappa_1}  
    + \phi_{j;\kappa_1} u^{[3]}_i P(t_0)_{\kappa_1 \gamma_1} v^{[3]}_{\gamma_1}\right]  
    \sigma_{v^{[3]}}^2   
    \label{eqn:dsttcov} \\
    &\quad + \frac{5}{12} \!\left[u^{[3]}_i u^{[3]}_j\right] \sigma_{v^{[3]}}(t_0)^6, \nonumber
\end{align}
where the $m$th-order R1-ODSTT elements are given by $u^{[m]}_i$ and the STM elements are given by $\phi_{i;j}$. This covariance propagation expression differs from the expression presented in \cite{boone2023directional} as it preserves the full STM in the first- and third-order cross-moments. \cref{eqn:sigma_r}-\ref{eqn:dsttcov} are the same as for R1-DSTTs with the relations $\psi^{[m]}_i = u^{[m]}_i$ and $R_i = v^{[m]}_i$ for all $m$. 

\section{Applications}
We will compare the approximation quality of the original R1-DSTT from \cite{boone2023directional} with the R1-ODSTT in the context of the two-body problem, circular restricted three-body problem, and aerocapture flight dynamics. 

\subsection{Dynamics}

\subsubsection{Two-Body Dynamics}

The equations of motion for the two-body problem in an inertial frame are
\begin{equation}
    \mathbf{F}(\mathbf{x}) = \begin{bmatrix}
        \dot{x} & \dot{y} & \dot{z} & -\frac{\mu x}{\|\mathbf{r}\|^3} & -\frac{\mu y}{\|\mathbf{r}\|^3} & -\frac{\mu y}{\|\mathbf{r}\|^3} 
    \end{bmatrix}^T,
\end{equation}
where the state vector is $\mathbf{x} = \left[x,y,z,\dot{x},\dot{y},\dot{z}\right]^T$, an overdot denotes a time derivative, $\mu$ is the standard gravitational parameter for the central body, and $\mathbf{r}=\left[ x,y,z\right]^T$ is the position vector from the central body to the satellite \cite{vallado_fundamentals_2001}. The results presented here are for a 300 km altitude circular LEO orbit propagated for three periods.

\subsubsection{Circular Restricted Three-Body Dynamics}

The equations of motion for the circular restricted three-body problem (CR3BP) are given in the synodic frame as
\begin{equation}
    \mathbf{F}(\mathbf{x}) = \begin{bmatrix}
        \dot{x} & \dot{y} & \dot{z} & 2\dot{y} + \frac{\partial \bar{U}}{\partial x} & 2\dot{x} + \frac{\partial \bar{U}}{\partial y} & \frac{\partial \bar{U}}{\partial z}
    \end{bmatrix}^T,
\end{equation}
where $\bar{U}(x,y,z)=\frac{1-\mu^*}{\|\mathbf{r}_1\|} + \frac{\mu^*}{\|\mathbf{r}_2\|}+\frac{x^2+y^2}{2}$ is the effective potential given by the reduced mass $\mu^* = \frac{m_2}{m_1+m_2}$ for the two primary bodies with masses $m_1$ and $m_2$ such that $m_1\geq m_2$ located along the x-axis at $[-\mu^*,0,0]$ and $[1-\mu^*,0,0]$ with respect to their common barycenter at the origin. The position of the satellite of interest with respect to the two primary bodies is given by $\mathbf{r}_1$ and $\mathbf{r}_2$ respectively \cite{Koon2000}. The reference orbit used in the following sections comes from the proposed NASA Gateway orbit \cite{Cunningham2023,NASA2019_NRHO}. Initial conditions for the orbit are
\begin{equation}
    \mu = 1.0/(81.30059 + 1.0), \quad x_0 = 1.022022, \quad z_0 = -0.182097, \quad \text{and} \quad \dot{y}_0 = -0.103256, 
\end{equation}
in nondimensional units with other initial coordinates equal to zero. This initial condition is at apolune of the orbit. The period of the orbit is 1.511111 nondimensional time units, where $2\pi$ time units is equivalent to the revolution period of the Earth-Moon system. 

\subsubsection{Aerocapture Dynamics}

The vehicle state is $\mathbf{x} = [r, \theta, \phi, V, \gamma, \psi, \ln(\rho)]^T$, where $r$ is the radial distance from the center of the planet, $\theta$ and $\phi$ are the latitude and longitude, $V$ is the planet-relative velocity, $\gamma$ is the flight path angle, $\psi$ is the heading angle, and $\ln(\rho)$ is the natural log of density. The 3DOF equations of motion for the spacecraft inside the atmosphere of a rotating planet are \cite{vihn_hypersonic_1980}:
\begin{align} \label{eqn:position}
	\dot{r}&=  V \sin \gamma, \\ 
	\dot{\theta}&=  \frac{V \cos \gamma \sin \psi}{r \cos \phi},\\
	\dot{\phi}&=  \frac{V \cos \gamma \cos \psi }{ r},\\
	\dot{V}&=  -D- g_r \sin \gamma -g_\phi \cos{\gamma}\cos{\phi}+\Omega^2 r \cos \phi (\sin \gamma \cos \phi - \cos \gamma \sin \phi \cos \psi),\\
	\dot{\gamma}&=\frac{1}{V}\left[L \cos \sigma+\left(V^2 / r-g_r\right) \cos \gamma+g_\phi \sin \gamma \cos \psi+2 \Omega V \cos \phi \sin \psi\right. \\
		&\left.+\Omega^2 r \cos \phi(\cos \gamma \cos \phi+\sin \gamma \cos \psi \sin \phi)\right] , \nonumber \\
	\dot{\psi}&=\frac{1}{V}\left[\frac{L \sin \sigma}{\cos \gamma}+\frac{V^2}{r} \cos \gamma \sin \psi \tan \phi+g_\phi \frac{\sin \psi}{\cos \gamma}-2 \Omega V(\tan \gamma \cos \psi \cos \phi-\sin \phi)\right. \\
		&\left.+\frac{\Omega^2 r}{\cos \gamma} \sin \psi \sin \phi \cos \phi\right], \qquad \text{and} \nonumber \\
 	\label{eqn:density}
	\dot{\ln(\rho)} &= -\frac{V \sin\gamma}{H},	
\end{align}
where $\sigma$ is the bank angle, $L$ and $D$ are the the lift and drag accelerations, $\Omega$ is the planet's constant angular velocity, $R_p$ is the planet's radius, and $H$ is the exponential atmosphere scale height.  To improve numerical stability, the nondimensionalization scheme developed by Lu was implemented \cite{lu_entry_2014}. A mass normalization factor $m_{ref}$ was also introduced, and density is normalized as $\rho^* = \rho / (m_{ref} / R_p^3)$. The reference mass was chosen such that natural log of density was nondimensionalized by $m_{ref} / R_p^3 = 20$. 

The equations to compute the latitudinal and longitudinal components of the gravity vector ($g_\theta$ and $g_\phi$) using $J_2$ are given in \cite{vihn_hypersonic_1980}. Lift and drag accelerations are computed as $L =  \frac{1}{2} \rho V^2 \frac{L}{D} \beta $ and $D =  \frac{1}{2} \rho V^2 \beta$, where $\frac{L}{D}=0.25$ is the lift-to-drag ratio and $\beta=145$ kg/m$^2$ is the ballistic coefficient. The natural logarithm of density is used as a state variable to include variations due to density change in the state. The density equation of motion was derived assuming an exponential atmosphere model $\rho = \rho_0 \exp{\frac{h_0 -(r - R_p)}{H}}$, where $\rho_0=6.40\mathrm{e}{-3}$ kg/m$^3$ is the reference density, $h_0=0$ km is the reference height, and $H=54.72$ km is the scale height.\footnote{The exponential atmosphere constants were found by fitting the data from UranusGRAM \cite{justh_uranus_2021}.}

The initial state is given in \cref{tab:init_state}; note that the position is given as altitude, and velocity and flight path angle are in the inertial frame. A constant bank angle of $\sigma=78^{\circ}$ is used. These values were chosen to be similar to those of the Flagship-class Uranus Orbiter Probe mission \cite{deshmukh_performance_2024}.

\begin{table}[h!]
\centering
\caption{Initial Aerocapture State.}
\label{tab:init_state}
\begin{tabular}{ccccccc}
    \hline
    $h$ [km] & $\theta$ [$^\circ$] &  $\phi$ [$^\circ$] &  $V$ [km/s] & $\gamma$ [$^\circ$] & $\psi$ [$^\circ$] & $\ln \rho$  [$\ln$(kg/m$^3$)]\\
    \hline
    1000 & 190.05 & -9.76 & 24.93 & -10.58 & 45 & -23.32\\ 
    \hline
\end{tabular}
\end{table}

\subsection{R1-ODSTT Approximation Accuracy}

\begin{figure}[htb]
     \centering
     \begin{subfigure}[b]{0.32\textwidth}
         \centering
         \includegraphics{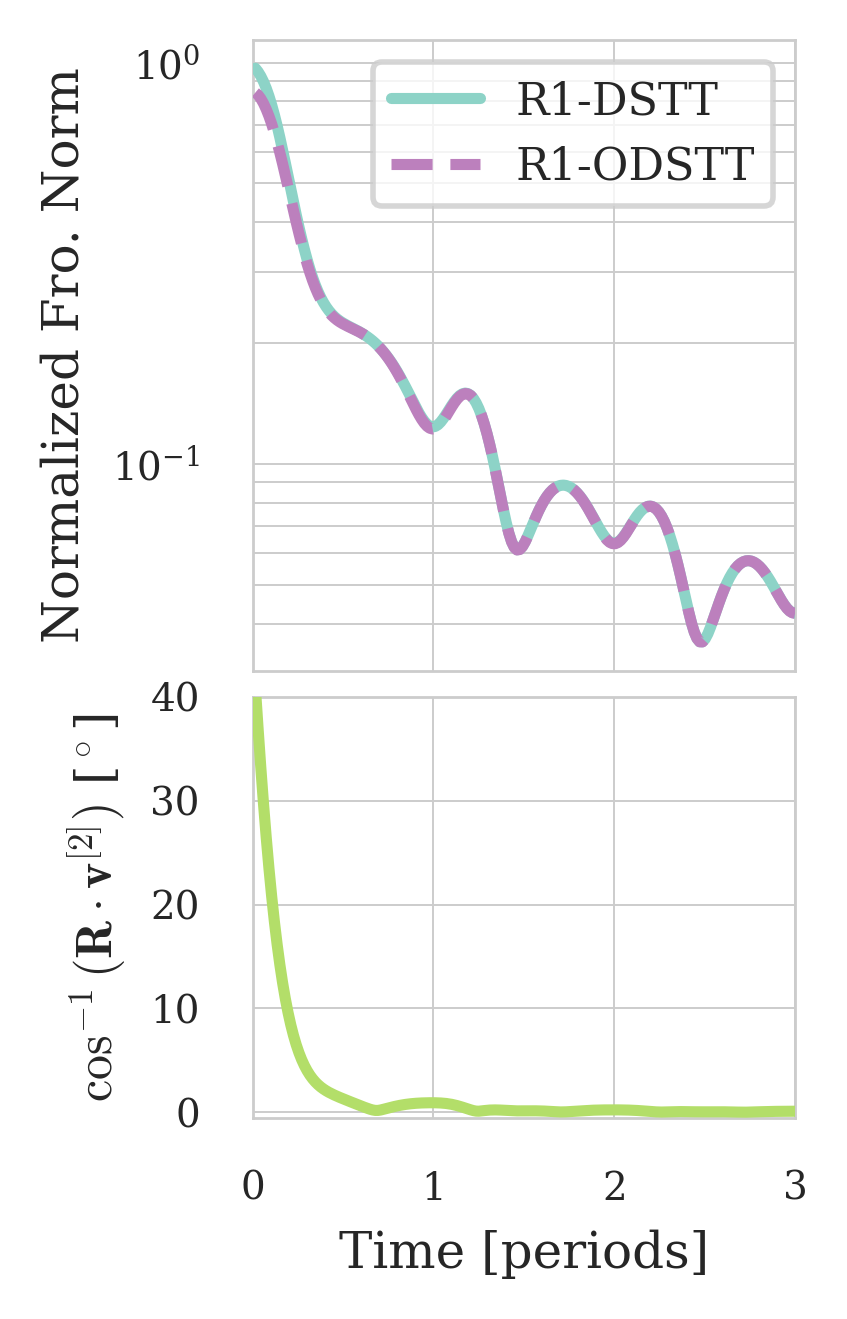}
         \caption{Two-Body Dynamics}
     \end{subfigure}
     \hfill
     \begin{subfigure}[b]{0.32\textwidth}
         \centering
         \includegraphics{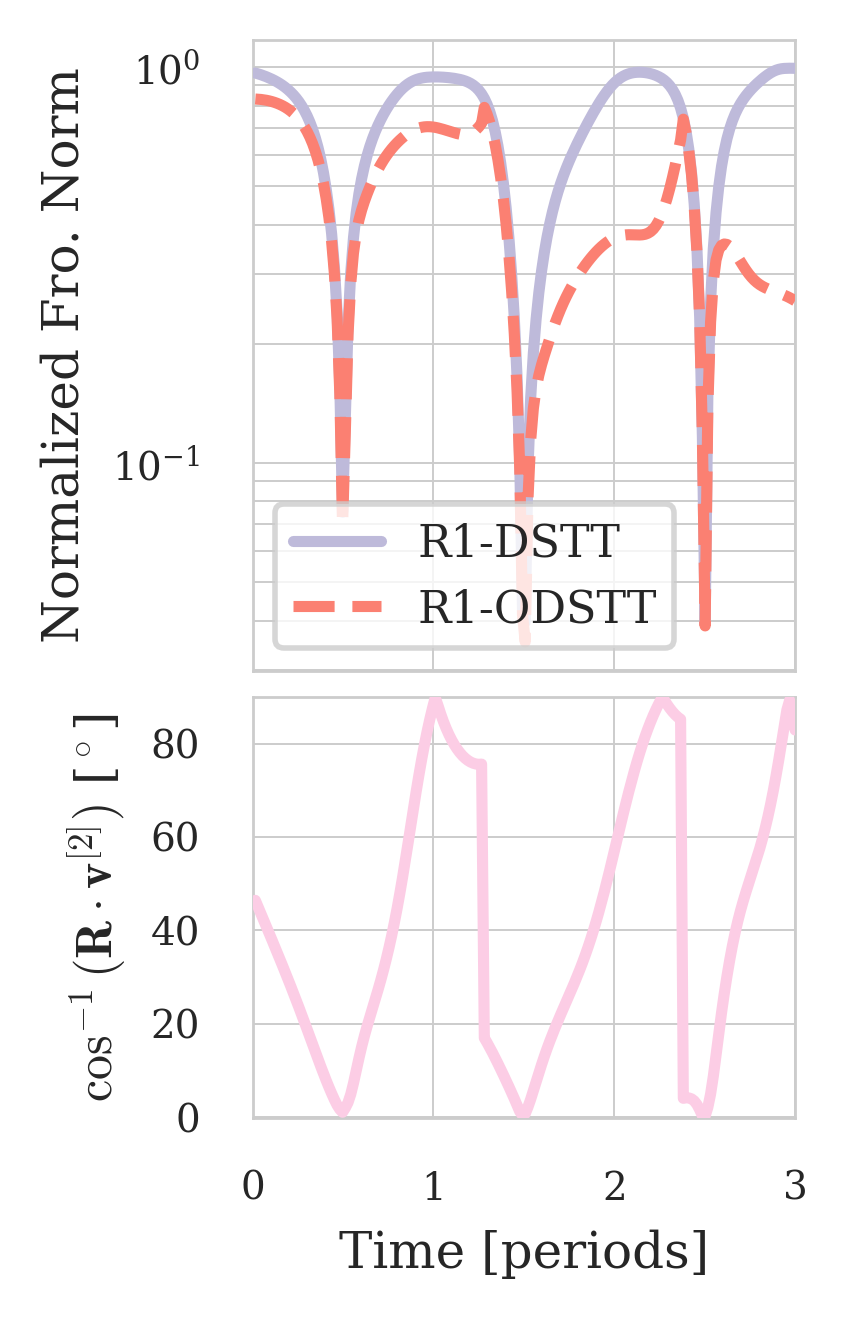}
         \caption{CR3BP}
     \end{subfigure}
     \hfill
     \begin{subfigure}[b]{0.32\textwidth}
         \centering
         \includegraphics{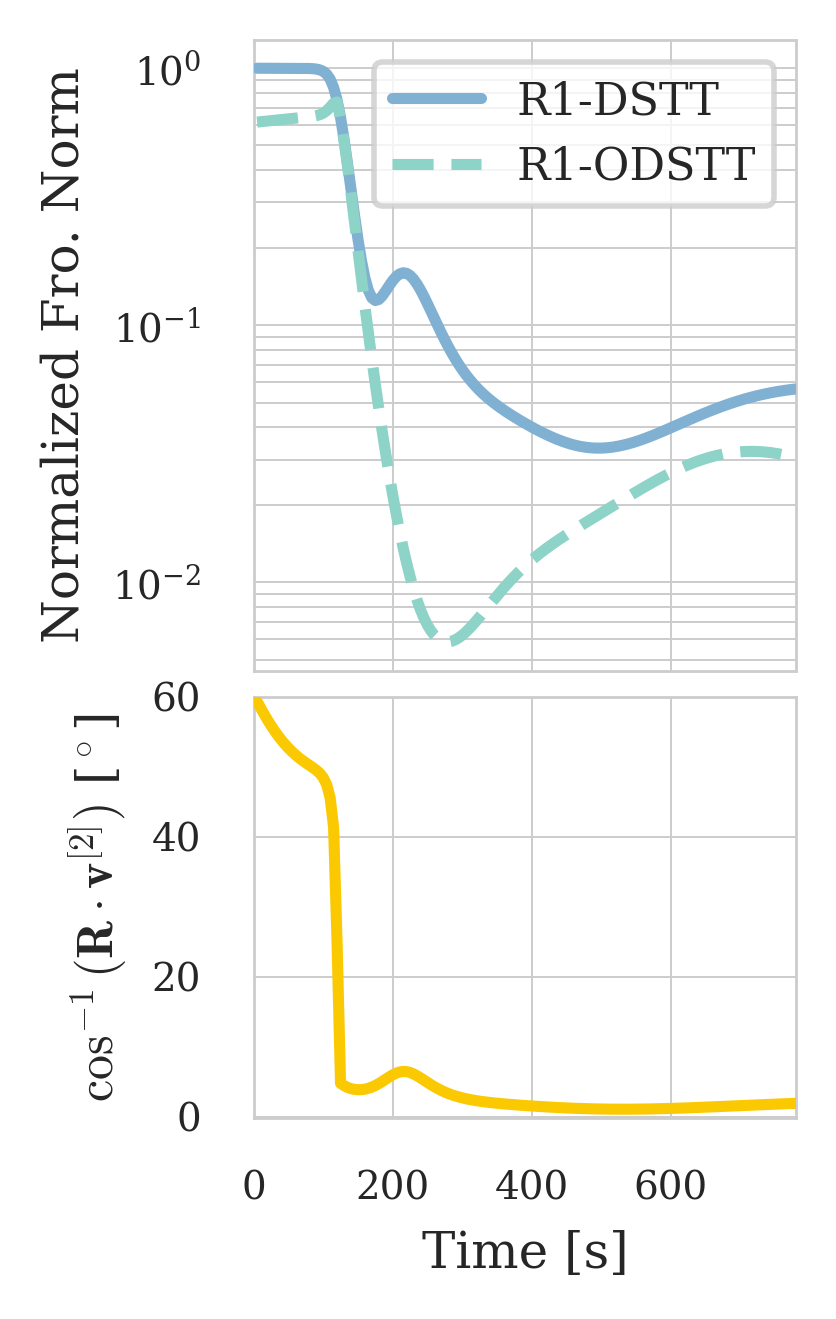}
         \caption{Aerocapture}
     \end{subfigure}
        \caption{Normalized Frobenius Norm Error.}
        \label{fig:frobenius_norm_error}
\end{figure}

The Frobenius norm of errors between the second-order STT and its R1-DSTT and R1-ODSTT approximations normalized by the second-order STT Frobenius norm are shown in \cref{fig:frobenius_norm_error} for all dynamics models. For the two-body dynamics, the R1-DSTTs and R1-ODSTTs perform similarly throughout the trajectory. However, for the more nonlinear CR3BP and aerocapture scenarios, the R1-ODSTTs outperform the R1-DSTTs. In addition, the aerocapture R1-ODSTTs are closer to the STTs near maximum dynamic pressure (around 250 s), which is a critical phase of flight for uncertainty propagation. This demonstrates the fact proven above: the R1-ODSTT approach results in a simplified STT with the most information retention in the Frobenius-norm sense. Although the R1-DSTTs and R1-ODSTTs have similar approximation error in the latter portion of the aerocapture trajectory, the propagation errors that occur when using R1-DSTTs near peak dynamic pressure can compound over time and degrade prediction accuracy later in the trajectory.

\cref{fig:frobenius_norm_error} also shows the angle between $\mathbf{R}$ and $\mathbf{v}^{[2]}$, $\cos^{-1}(\mathbf{R} \cdot \mathbf{v}^{[2]})$, which is $0^\circ$ when the R1-DSTT $\mathbf{R}$ and R1-ODSTT $\mathbf{v}^{[2]}$ input directions are colinear and $90^\circ$ when they are orthogonal. For the two-body problem, where the Frobenius norm error is similar between the R1-ODSTTs and R1-DSTTs, the $\mathbf{R}$ and $\mathbf{v}^{[2]}$ directions are near each other throughout the entire trajectory. For this case, the dominant right singular subspace of the state transition matrix is a good approximation for the optimal $\mathbf{v}^{[2]}$ direction found with the R1-ODSTT approach. This indicates that the direction of maximum stretching for a linearly propagated perturbation is similar to the direction that preserves the most higher-order information from the second-order STT, which is consistent with Boone and McMahon's findings \cite{boone2023directional}. 

However, for the more nonlinear CR3BP and aerocapture dynamics, the $\mathbf{R}$ and $\mathbf{v}^{[2]}$ directions are more dissimilar. In the CR3BP, the R1-ODSTTs and R1-DSTTs both have 10\% approximation error near perilune as the normalized Frobenius norm error magnitude dips, and at those same points the R1-ODSTT and R1-DSTT input directions are similar. However, away from perilune, where both approximations are less effective but the R1-ODSTT is a better approximation of the STT, there is a clear difference between the $\mathbf{R}$ and $\mathbf{v}^{[2]}$ directions. For the CR3BP, the difference in $\mathbf{R}$ and $\mathbf{v}^{[2]}$ directions is well correlated with the degree to which the R1-ODSTTs outperform the R1-DSTTs. In contrast, the $\mathbf{R}$ and $\mathbf{v}^{[2]}$ directions become closer throughout the trajectory in the aerocapture scenario, but the R1-ODSTT still outperforms the R1-DSTT in terms of Frobenius norm error. This suggests that even a small divergence between the $\mathbf{R}$ and $\mathbf{v}^{[2]}$ directions can lead to a considerable difference in approximation error.

While the original R1-DSTT approach yields a near-optimal input direction and R1-DSTT in the Frobenius-norm sense for the two-body problem, this does not hold true for more nonlinear scenarios like the CR3BP and aerocapture. For such scenarios, the R1-ODSTTs retain more of the higher-order STT information than the R1-DSTTs, especially in critical nonlinear regions like near maximum dynamic pressure in the aerocapture problem and away from perilune in the CR3BP.

\subsection{R1-ODSTTs Applied to Aerocapture Uncertainty Quantification}

The second and third-order R1-ODSTTs and R1-DSTTs are compared with second-and third-order STTs for propagation of a Gaussian distribution with $\delta \mathbf{m}(t_0) = 0$ through the nonlinear aerocapture dynamics. The mean and covariance are propagated with STTs following \cref{eqn:stt_mean,eqn:sttcov} and with R1-DSTTs (preserving the full STM) following \cref{eqn:dsttmean,eqn:dsttcov}. The R1-DSTT expressions can be used for second-order propagation by setting $\mathbf{u}^{[3]}=0$. All states have an initial nondimensional variance of $5\times 10^{-7}$ such that the state distribution remains near-Gaussian. The initial dimensional covariance matrix elements are given in \cref{tab:init_cov}.

\begin{table}[h!]
\centering 
\caption{Initial Covariance Matrix Elements.}
\label{tab:init_cov}
\begin{tabular}{ccccccc}
    \hline
    $\sigma_r$ [km] & $\sigma_\theta$ [$^\circ$] &  $\sigma_\phi$ [$^\circ$] &  $\sigma_V$ [km/s] & $\sigma_\gamma$ [$^\circ$] & $\sigma_\psi$ [$^\circ$] & $\sigma_{\ln\rho}$ [$\ln$(kg/m$^3$)] \\
    \hline
    18.07 & 4.05e${-2}$ & 4.05e${-2}$ & 0.12 & 4.05e${-2}$ & 4.05e${-2}$ & 1.41e${-2}$\\
\end{tabular}
\end{table}

\begin{figure}[htb]
     \centering
     \begin{subfigure}[b]{0.47\textwidth}
         \centering
         \includegraphics{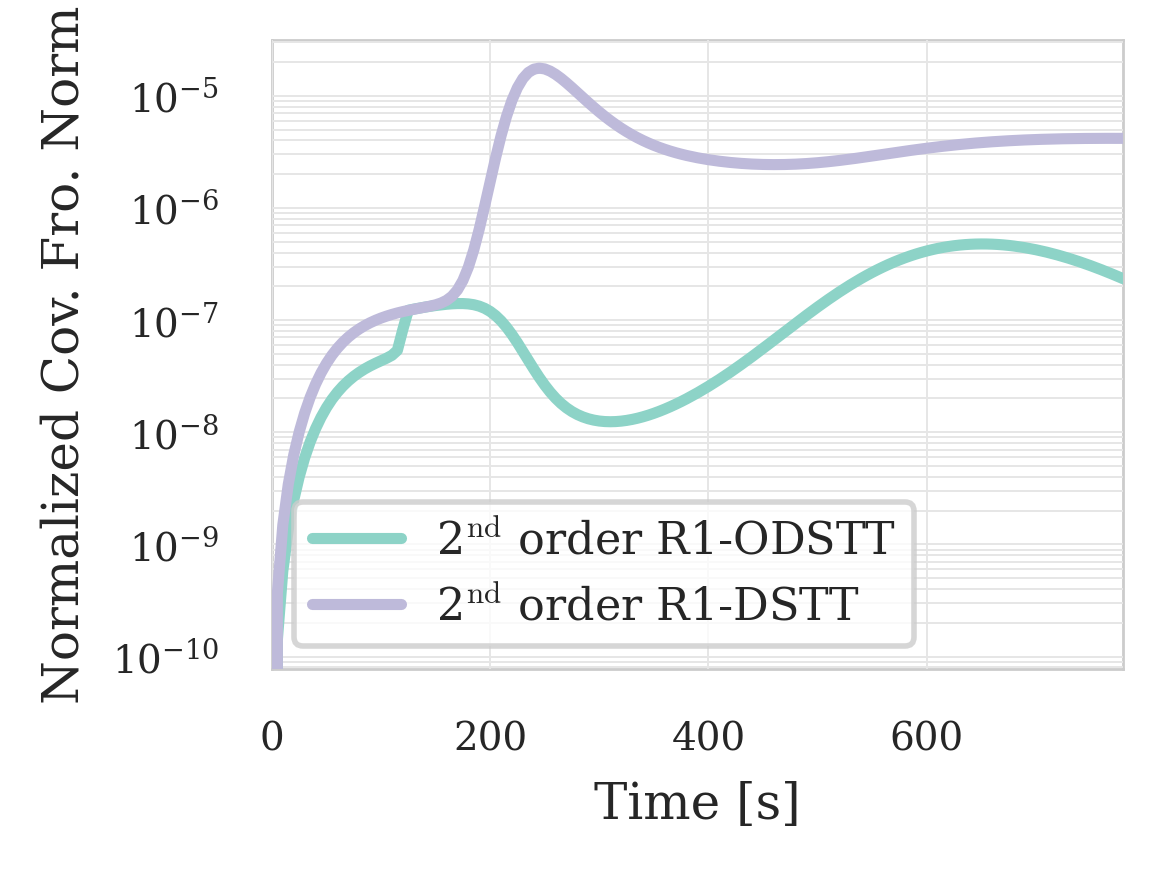}
         \caption{Second-Order}
         \label{fig:fro_norm_cov_2}
     \end{subfigure}
     \hfill
     \begin{subfigure}[b]{0.47\textwidth}
         \centering
         \includegraphics{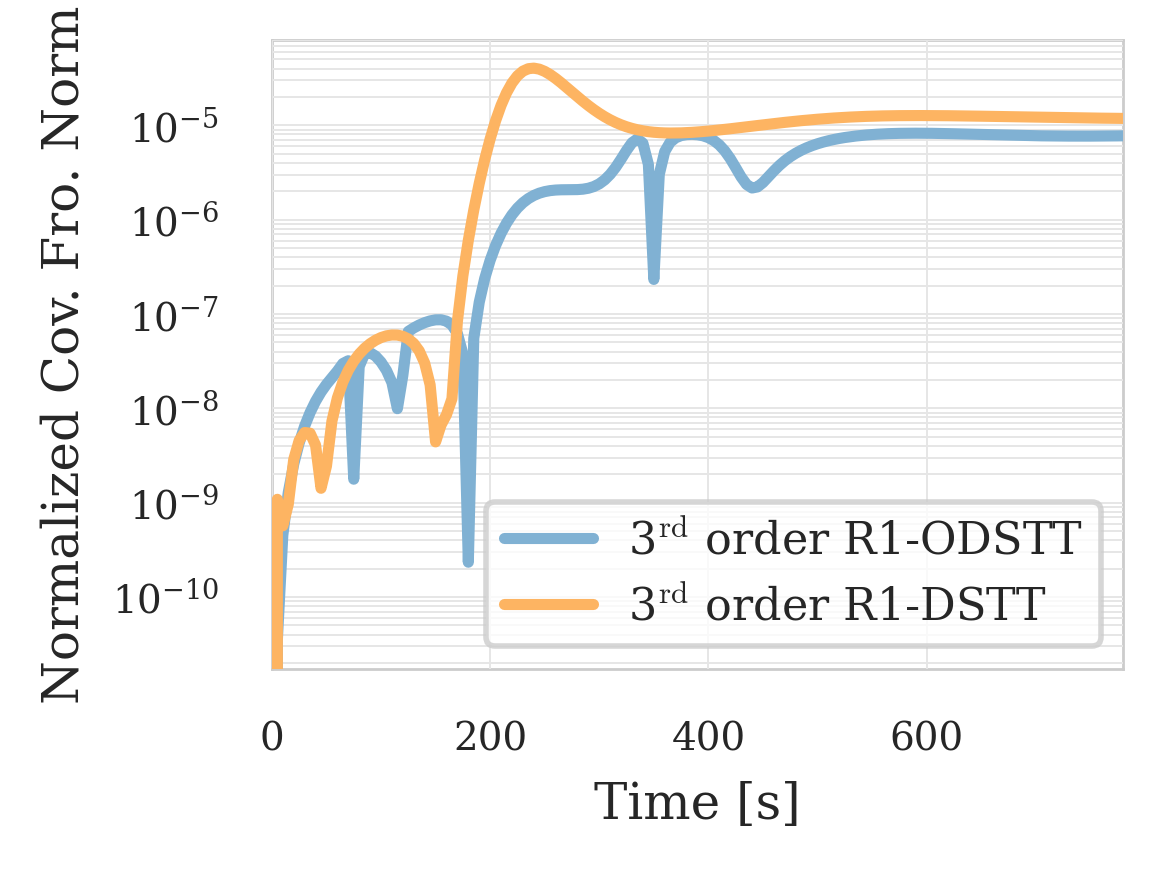}
         \caption{Third-Order}
         \label{fig:fro_norm_cov_3}
     \end{subfigure}
    \caption{Normalized Frobenius Norm Error of Nondimensional R1-DSTT Propagated Covariance.}
    \label{fig:fro_norm_cov}
\end{figure}

\cref{fig:fro_norm_cov} shows the normalized Frobenius norm difference between a R1-DSTT propagated covariance, $P_\text{R1-DSTT}$, and the STT propagated covariance, $P_\text{STT}$, for a given order:
\begin{equation}
    \frac{\| P_\text{STT} - P_\text{R1-DSTT}\|_F}{\|P_\text{STT}\|_F}.
\end{equation}

The second-order R1-ODSTT-propagated covariances have a lower normalized Frobenius norm error than the second-order R1-DSTT-propagated covariances, indicating that the ODSTT-propagated distribution is closer to the second-order STT results. This shows that the second-order R1-ODSTTs are not only better approximations of the second-order STTs in the Frobenius-norm sense, but are more effective at second-order moment propagation for nonlinear problems. However, the third-order R1-DSTT-propagated covariance has two regions (approximately between 25 and 75 seconds and between 120 and 175 seconds) where its normalized covariance Frobenius norm error is lower than the R1-ODSTT error of the same order. These regions could be due to the R1-DSTT directions for third-order better capturing nonlinearity in the covariance dynamics than the R1-ODSTT directions. Over certain time regimes, the R1-DSTT third-order direction preserves more information from the covariance matrix than the R1-ODSTT third-order direction for this homoscedastic initial distribution. While the ODSTT is always optimal in the Frobenius norm sense, it may not always be optimal for covariance propagation. Additionally, the magnitudes of the third-order R1-DSTT and R1-ODSTT normalized Frobenius norm error are higher than the second-order R1-DSTT and R1-ODSTT errors in the latter three-quarters of the trajectory. This does not indicate that the third-order R1-DSTTs and R1-ODSTTs are less accurate than the second-order R1-DSTTs and R1-ODSTTs, but rather that the third-order R1-DSTT covariances are further from the third-order STT covariances than the second-order R1-DSTT covariances are from the second-order STT covariances. 

\subsection{Validation of R1-ODSTT Error Bound}
In the past two sections, we have shown that the R1-ODSTTs can improve performance over the R1-DSTTs in terms of normalized Frobenius norm error and moment propagation. As proven in \cref{sec:ODSTTs}, the largest possible error incurred when propagating a vector on the unit ball using a second-order R1-ODSTTs instead of STTs is smaller than the Frobenius norm of the error tensor. To demonstrate this bound, 1000 initial perturbations are sampled from a standard multivariate normal distribution $\delta \mathbf{x}_0 \sim \mathcal{N}(0_{n\times 1}, \mathbf{I_{n\times n}})$ and normalized to unit vectors.

\begin{figure}[htb]
     \centering
     \includegraphics{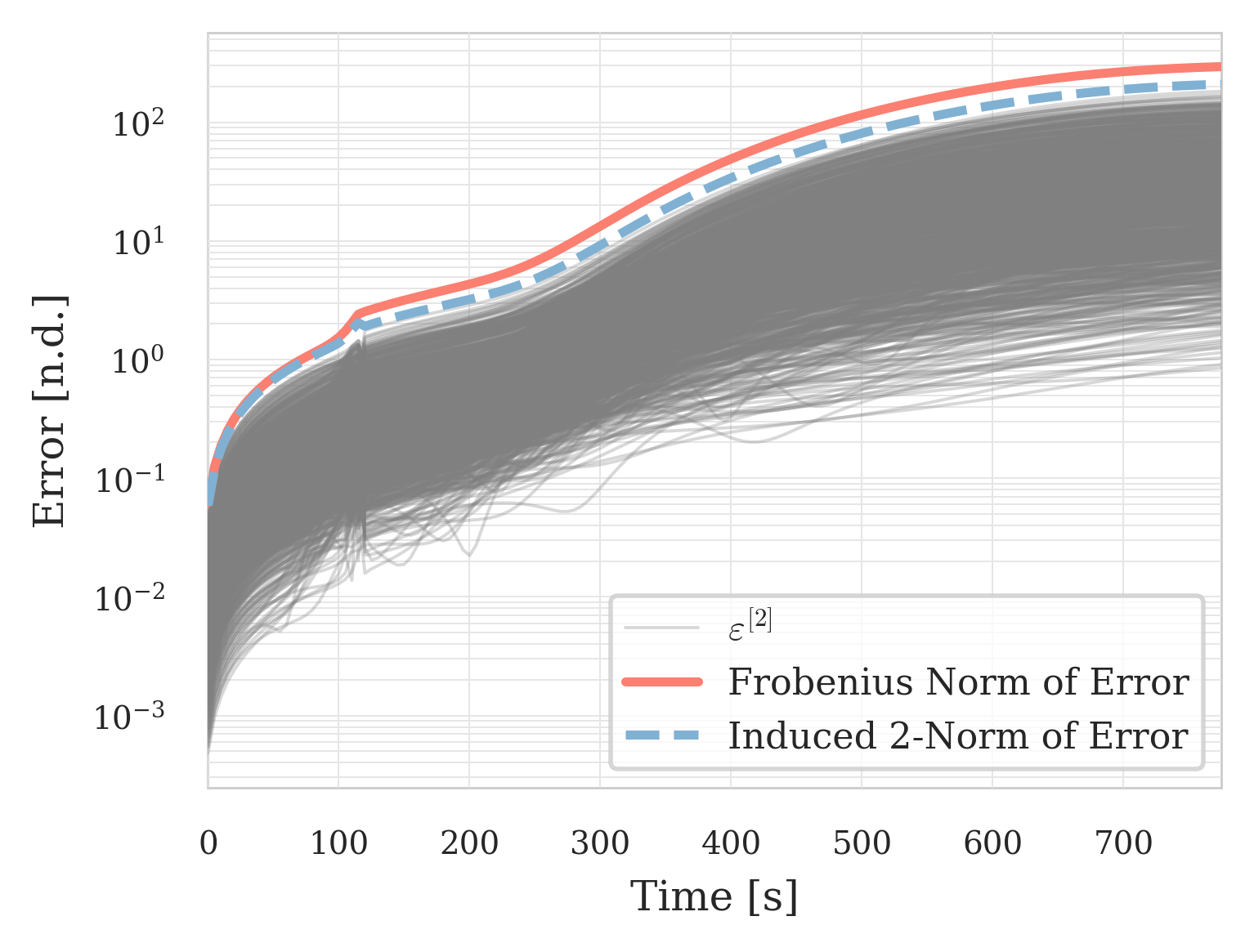}
     \caption{Norm of perturbation propagation error using R1-ODSTTs.}
     \label{fig:error_odstt}
\end{figure}

\cref{fig:error_odstt} shows the norm of the error between the time history of each perturbation propagated with the second-order R1-ODSTT and second-order STT, 
\begin{equation}
\varepsilon^{[2]}(t_k) = \left| \left| \boldsymbol{\Phi}^{[2]} \otimes \delta \mathbf{x}(t_0) \otimes \delta \mathbf{x}(t_0) - \mathbf{u}\otimes (\mathbf{v} \cdot \delta \mathbf{x}(t_0))\otimes (\mathbf{v} \cdot \delta \mathbf{x}(t_0))\right| \right|,
\end{equation}
where $\boldsymbol{\Phi}^{[2]}$ is the second-order STT, and all mappings are from $t_0$ to $t_k$. These quantities are not the true $\delta \mathbf{x}(t_k)$, but rather twice the second-order contribution to $\delta \mathbf{x}(t_k)$. Given that both propagation methods use the STM for first-order dynamics, focusing on this term isolates the second-order contribution to the overall error. The individual errors are compared with the Frobenius norm of the error (the left side of \cref{eqn:fro_norm_of_error}) and the induced 2-norm of the error (the right side of \cref{eqn:fro_norm_of_error}). As proven above, all errors are less than the theoretical bound, and the induced 2-norm of the error tensor is less than or equal to the Frobenius norm of the error tensor. 

\section{Conclusion}

An optimal rank-1 approximation of state transition tensors was developed as an efficient alternative to state transition tensors for nonlinear uncertainty quantification. While previous directional state transition tensors used the dominant right singular subspace of the state transition matrix to construct a reduced-dimension representation of the state transition tensors, optimal directional state transition tensors are constructed to maximize the information retained in a rank-1 approximation of the state transition tensors in the Frobenius-norm sense. The optimal rank-1 directional state transition tensor is found by solving a tensor z-eigenpair problem of the ``square'' of the state transition tensor. This construct leads to increased approximation accuracy of the state transition tensors and improved Gaussian moment propagation for nonlinear flight scenarios like aerocapture. However, for more linear dynamics such as the two-body problem, the original directional state transition tensor approach is near-optimal, and little improvement is gained from the optimal approach. The ODSTTs are guaranteed to be more accurate in the Frobenius norm sense than the DSTTs, and thus are a strong first choice when constructing rank-1 DSTTs over the original DSTT approach. Yet, finding the optimal input and output directions requires solving the more computationally demanding tensor eigenproblem, rather than the matrix eigenproblem required for the original DSTT approach. This tradeoff in computational expense when computing the DSTTs may not always be worthwhile if the optimal $\mathbf{v}$ direction is near the $\mathbf{R}$ direction, as little performance improvement can be achieved. While rank-$k$ ($k \leq n$) DSTTs can easily be constructed, the ODSTT approach can be extended find the optimal rank-$k$ approximation of the state transition tensors. This is left as future work.

\section*{Funding Sources}
This work was supported by NASA Space Grant Technology Research Fellowship grant number 80NSSC23K1227. 

\section*{Acknowledgments}
J. Kulik would like to thank Max Ruth for his insights on low-rank approximation.

\bibliography{references}

\end{document}